\documentclass[twocolumn]{revtex4}
\usepackage{mathrsfs}
\usepackage{graphicx}
\usepackage{amsmath}
\usepackage{booktabs}
\usepackage{amssymb}
\usepackage{epsfig}
\usepackage{times}
\usepackage{color}
\usepackage{float}

\usepackage{subfigure}

\begin{document}
\title{Zero-Determinant Strategies in the Iterated Public Goods Game}

\author{Liming Pan}
\author{Dong Hao}
\author{Zhihai Rong}
\author{Tao Zhou}
\email{zhutou@ustc.edu}

\affiliation{
Web Sciences Center, University of Electronic Science and Technology of China, Chengdu 611731, People's Republic of China
}

\begin{abstract}
Recently, Press and Dyson have proposed a new class of probabilistic
and conditional strategies for the two-player iterated Prisoner's
Dilemma, so-called zero-determinant strategies. A player adopting
zero-determinant strategies is able to pin the expected payoff of
the opponents or to enforce a linear relationship between his own
payoff and the opponents' payoff, in a unilateral way. This paper
considers zero-determinant strategies in the iterated public goods
game, a representative multi-player evolutionary game where in each
round each player will choose whether or not put his tokens into a
public pot, and the tokens in this pot are multiplied by a factor
larger than one and then evenly divided among all players. The
analytical and numerical results exhibit a similar yet different
scenario to the case of two-player games: (i) with small number of
players or a small multiplication factor, a player is able to
unilaterally pin the expected total payoff of all other players;
(ii) a player is able to set the ratio between his payoff and the
total payoff of all other players, but this ratio is limited by an
upper bound if the multiplication factor exceeds a threshold that
depends on the number of players.
\end{abstract}


\maketitle

\section{Introduction}
Iterated games have long been exemplary models for the emergence of
cooperations in socioeconomic and biological systems
\cite{Axelrod1981,Axelrod1984,Axelrod1988,Nowak1993,Nowak2006,Kendall2007}.
Learned from these studies, the most significant lesson is that in
the long term, selfish behavior will hurt you as much as your
opponents. Therefore, from both scientific and moral perspectives,
ants and us all live in a reassuring world: altruists will
eventually dominate a reasonable population. Very recently, however,
Press and Dyson \cite{Press2012} have shattered this well-accepted
scenario by introducing a new class of probabilistic memory-one
strategies for the two-player iterated Prisoner's Dilemma (IPD),
so-called zero-determinant (ZD) strategies. Via ZD strategies, a
player can unilaterally pin his opponents' expected payoff or extort
his opponents by enforcing a linear relationship between his own
payoff and the opponents' payoff. In a word, egotists could become
more powerful and harmful if they know mathematics. Though being
challenged by the evolutionary stability
\cite{Adami2013,Hilbe2013,Stewart2013}, studies on ZD strategies as
a whole
\cite{Press2012,Adami2013,Hilbe2013,Stewart2013,Akin2012,Chen2013,Hilbe2013b,Daoud2014,Szolnoki2014}
will dramatically change our understanding on iterated games
\cite{Stewart2012,Hayes2013}. Indeed, knowing the existence of ZD
strategies has already changed the game.

ZD strategies in IPD can be naturally extended to other iterated
two-player games \cite{Roemheld2013}, which are still uncultivated
lands for scientists. Instead, we turn our attention to the iterated
multi-player games and try to answer a blazing question: could a
single ZD player in a group of considerable number of players
unilaterally pin the expected total payoff of all other players and
extort them? This paper focuses on a notable representative of
multi-player games, the public goods game (PGG)
\cite{Hardin1968,Kagel1995}. In the simplest $N$-player PGG, each
player chooses whether or not contribute a unit of cost into a
public pot. The total contribution in the public pot will be
multiplied by a factor $r$ ($1<r<N$) and then be evenly divided
among all $N$ players, regardless whether they have contributed or
not. As a simple but rich model, the PGG arises a question why and
when a player is willing to contribute against the obvious Nash
equilibrium at zero \cite{Fehr2003}, which is critical for the
understanding, predicting and intervening of many important issues
ranging from micro-organism behaviors
\cite{Cordero2012,Bachmann2013} to global warming
\cite{Milinski2006,Milinski2008,Tavonia2011,Santos2011}. Among a
couple of candidates
\cite{SigmundPNAS01,Hauert2002,SzaboPRL2002,Santos2008,RongPRE2010,Apicella2012,Perc2013},
the repeated interactions may be a relevant mechanism to the above
question, since reputation, trustiness, reward and punishment can
then play a role \cite{Fehr2000,Milinski2002}. We thus study the
iterated public goods game (IPGG, also named as repeated public
goods game in the literatures) where the same players in a group
play a series of stage games.

It is found by surprise that the ZD strategies still exist for a
group with many players in IPGG, namely a single player can pin the
total payoff of all others or extort them in a unilateral way.
However, different from the observations in IPD, there exists some
unreported restrictive conditions related to the group size and
multiplication factor, which determine the feasibility to pin the
total payoff of all other players and the upper bound of
extortionate ratio.

\begin{figure*} 
\subfigure[ ] { \label{fig1:a}
\includegraphics[width=1\columnwidth]{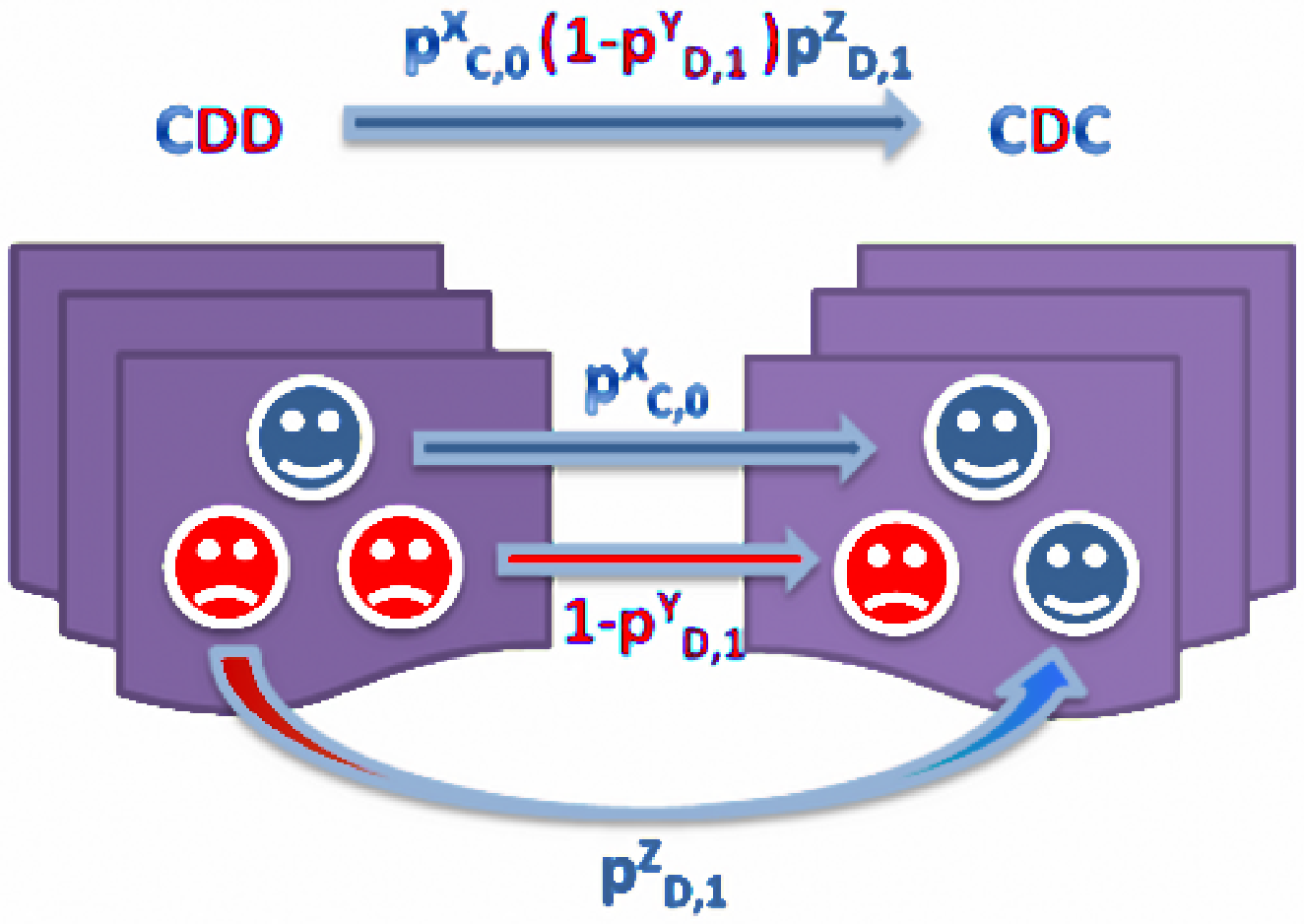}
} \subfigure[ ] { \label{fig1:b}
\includegraphics[width=0.85\columnwidth]{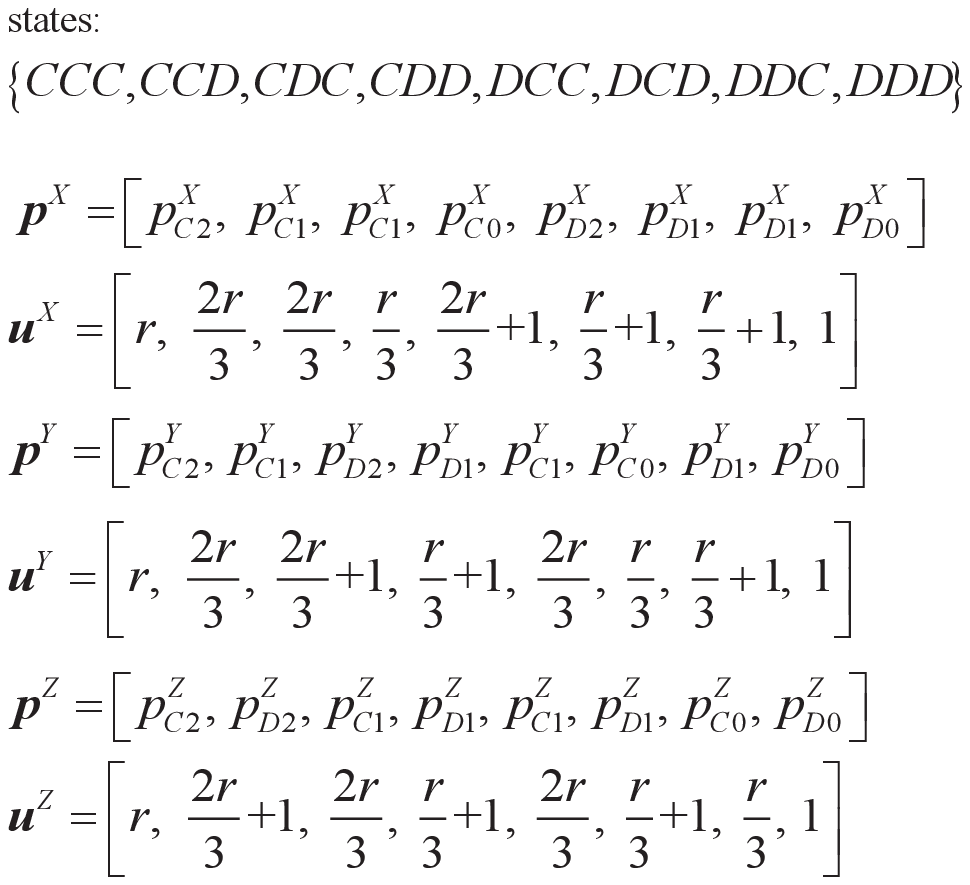}}

\subfigure[ ] { \label{fig1:c}
\includegraphics[width=2\columnwidth]{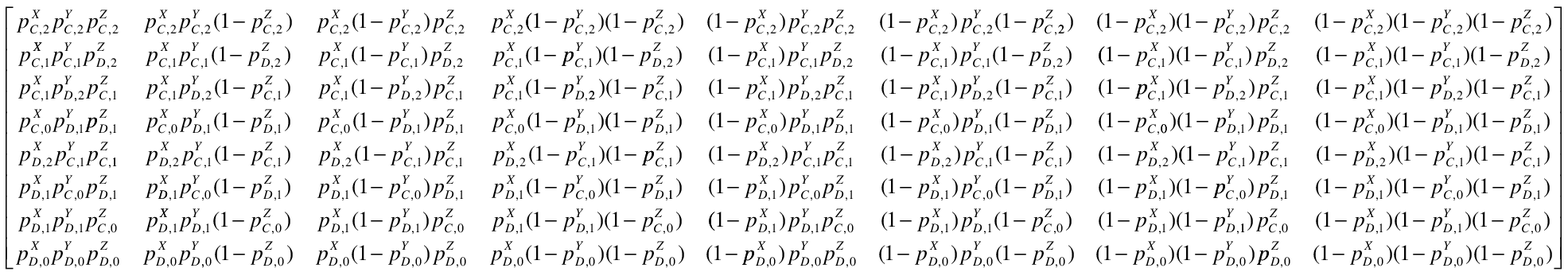}}

\subfigure[ ] { \label{fig1:d}
\includegraphics[width=2\columnwidth]{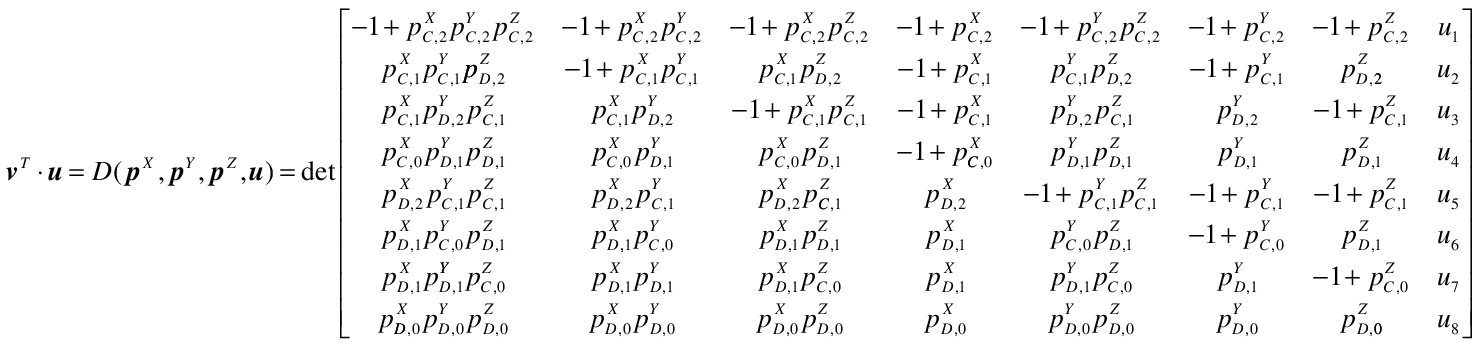}

}
\caption{Illustration of the general three-player game. (a) For a
previous state $CDD$, the conditional probabilities that the players
$X$, $Y$ and $Z$ select $C$ in the current round are $p^X_{C,0}$,
$p^Y_{D,1}$ and $p^Z_{D,1}$, respectively. Therefore, the
probability from the previous state $CDD$ to the current state $CDC$
is $p^X_{C,0}(1-p^Y_{D,1})p^Z_{D,1}$. (b) shows the $2^3=8$
different states, the strategy vectors and payoff vectors of the
three players. According to Eq. (\ref{eq:markov}) and Eq.
(\ref{eq:smallmk}), the Markov transition matrix {\bf{M}} for the
present case is shown in (c). After some elementary column
operations on matrix $\bf{M-I}$, the dot product of an arbitrary
vector $\bf{u}$ with the stationary vector $\bf{v}$ is equal to the
determinant $\det({{\bf{p}}^X},{{\bf{p}}^Y},{{\bf{p}}^Z},{\bf{u}} )$
as shown in (d), where ${\bf{\tilde{p}}}^X $, ${\bf{\tilde{p}}}^Y $
and ${\bf{\tilde{p}}}^Z $ lie in the forth, sixth and seventh
columns which are only controlled by the players $X$, $Y$ and $Z$,
respectively. } \label{ThreePlayer}
\end{figure*}
%
%
%

\section{ZD Strategies in Multi-Player Games}
Consider an $N$-player iterated game, which consists of a series of
repetitions of a same stage game of $N$ players. Press and Dyson
\cite{Press2012} proved the theorem that, in such an iterated game,
if the stage games are identically repeated infinite times, a
long-memory player will have no advantage over a short-memory
player. Without loss of generality, it is suffice to derive players'
strategy assuming they have only memory-one. Thus, which action a
player will take in the current round depends on the outcome of the
previous round. For an arbitrary player $X\in\{1,2,\cdots,N \}$, a
(mixed) strategy ${\bf{p}}^X$ is a vector that consists of
conditional probabilities for cooperating with respect to every
possible outcome. Since we consider a general $N$-player game and
every player may choose cooperation ($C$) or defection ($D$), there
are $2^N$ possible outcomes for each round. For player $X$, his
memory-one strategy can be represented by a $2^N$-dimensional
vector:
\begin{equation}\label{eq:p1}
\begin{split}
{\bf{p}}^X=& \left[p^X_{1}, \ \cdots\ \,p^X_{i}, \ \cdots\
\,p^X_{2^N} \right]^{T},
\end{split}
\end{equation}
where $p^X_{i}$ stands for the conditional probability that $X$ will
cooperate in the current round, given the outcome of the previous
round. Here $i{\small{=}}1,2,\cdots ,2^N$ is the index of possible
outcomes in each round. Figure 1 shows an example for an
$N{\small{=}}3$ game, in which the players are $X$, $Y$ and $Z$, and
the possible outcomes are $\{CCC,CCD,CDC,CDD,DCC,DCD,DDC,DDD \}$.

In many multi-player games such as public goods
game \cite{Hardin1968,Kagel1995}, N-player snowdrift game \cite{ZhengEPL07},
N-player stag-hunt game \cite{PachecoPRSB09} and collective-risk social dilemma \cite{Milinski2008},
whether a specific opponent chooses to
cooperate is less meaningful, instead, it is crucial for a player to
know how many his opponents cooperate. In such a scenario, if a
player $X$'s previous move is $C$ and the number of cooperators
among the opponents in the previous round is {\small{$n\in \{
0,1,\cdots,N-1 \} $}}, the probabilities for him to cooperate in the
current round are denoted as $p^X_{C,n}$. Similarly, if his previous
move is $D$ and the number of cooperating opponents is $n$, the
probability to cooperate is $p^X_{D,n}$. Therefore, the original
strategy vector in Eq.~(\ref{eq:p1}) can be refined to a
$2^N$-dimensional vector with $2N$ independent variables as:
\begin{equation}\label{eq:p2}
\begin{split}
{\bf{p}}^X=& [ p^X_{C,0},\ \cdots\ ,\underbrace{p^X_{C,n},
\cdots\,p^X_{C,n}}_{\binom{N-1}{n}\ terms}\ ,\cdots\ ,p^X_{C,N-1},\\
&p^X_{D,0},\ \cdots\ ,\underbrace{p^X_{D,n},
\cdots\,p^X_{D,n}}_{\binom{N-1}{n}\ terms}\ , \cdots,\ p^X_{D,N-1}
]^{T}.
\end{split}
\end{equation}
Figure~\ref{ThreePlayer}(b) gives an example of the strategy vectors
for the three-player case.

Starting at an initial outcome, the $N$ players' strategy profile
determines a stochastic process. Since these are memory-one
strategies, the corresponding stochastic process can be
characterized by a Markov chain. Each possible outcome of the repeated
games can be maintained by a state in this Markov chain model. Under
this model, the state transition rules are joint probabilities
calculated from the $N$ players' probabilistic strategies. Denoting
the corresponding transition matrix as:
\begin{equation}\label{eq:transmatrix}
\begin{split}
{\bf{M}} =\left[  M_{ij} \right]_{{2^N}\times{2^N}},
\end{split}
\end{equation}
where the element $M_{ij}$ is a one-step transition probability of moving from state $i$ to state $j$. It is essentially a joint probability that can be calculated as:
\begin{equation}\label{eq:markov}
M_{ij}=\prod\nolimits_{X=1}^{N}m^X,
\end{equation}
where $X$ runs over all players, and
\begin{equation}\label{eq:smallmk}
m^X  = \left\{ \begin{array}{l}
 (p_{C,n(i)}^X )^{h_j^X } (1 - p_{C,n(i)}^X )^{1 - h_j^X } ,{\textrm{ if $X$ takes $C$ in state $i$}}; \\
 (p_{D,n(i)}^X )^{h_j^X } (1 - p_{D,n(i)}^X )^{1 - h_j^X } ,{\textrm{if $X$ takes $D$ in state $i$}}. \\
 \end{array} \right.
\end{equation}
Here $n(i)$ is the number of cooperators among $X$'s opponents in
state $i$. $h_j^X$ is an indicator, a binary variable determined by
player $X$'s action in state $j$. Conventionally, if player $X$'s
action in state $j$ is $C$, then $h_j^X=1$; otherwise, $h_j^X=0$.
Figure 1(c) shows $\bf{M}$ for the general three-player game. It can be
easily checked that the sum of each row equals $1$.

In Eq. (\ref{eq:markov}) and Eq. (\ref{eq:smallmk}), the transition
probabilities are dependent on all the $N$ players' strategies, reflecting the complexity of the multi-player games.
However, the approach proposed by Press and Dyson \cite{Press2012}
allow us to derive a class of strategies succinctly, but profoundly.
Define a matrix ${\bf{M}}'={\bf{M - I}}$, where ${\bf{I}}$ is the
unit diagonal matrix. After some elementary column operations on
this matrix, the joint probabilities will be finely separated,
leaving one column solely controlled under player $X$'s strategy but
not dependent on other players anymore. This column
$\tilde{\bf{p}}^X$ is shown as follows:
\begin{equation}\label{eq:px}
\begin{split}
{\tilde{\bf{p}}^X}=&[-1+p^X_{C,0},\ \cdots\
,\underbrace{-1+p^X_{C,n},  \cdots\ ,-1+p^X_{C,n}}_{\binom{N-1}{n}\
terms}\ , \cdots\ \\&-1+p^X_{C,N-1},\ p^X_{D,0},\ \cdots\ ,
\underbrace{p^X_{D,n},  \cdots\ ,p^X_{D,n}}_{\binom{N-1}{n}\ terms}\
, \\& \cdots\ ,p^X_{D,N-1}]^{T}.
\end{split}
\end{equation}
In Eq. (\ref{eq:px}), all the probabilities depend only on the
elements in Eq. (\ref{eq:p2}), which indicates that
${\bf{\tilde{p}}}^X$ is unilaterally controlled by player $X$. Note
that ${\bf{\tilde{p}}}^X$ is a $2^N$-dimensional vector, and the
elements $-1+p_{C,n}^X$ and $p_{D,n}^X$ each appears
$\binom{N-1}{n}$ times. Figure \ref{ThreePlayer}(d) gives an
example of the unilateral control for the general three-player game. We can
see that the forth, sixth and seventh columns in this matrix only
involve the strategies of players $X$, $Y$ and $Z$, respectively.

If the transition matrix $\bf{M}$ is regular, i.e., the Markov chain
is irreducible and aperiodic, it will be ensured that there exists a
stationary probability vector ${\bf{v}}$, such that
\begin{eqnarray}\label{stationary}
\begin{array}{l}
{{\bf{v}}^T}\cdot{\bf{M}}= {{\bf{v}}^T}.
\end{array}
\end{eqnarray}
The stationary vector ${\bf{v}}$ is the very eigenvector
corresponding to the eigenvalue ${\bf{1}}$ of {\bf{M}}. Press and
Dyson ~\cite{Press2012} prove that, there is a proportional
relationship between the stationary vector $\bf{v}$ and each row in
the adjugate matrix ${\mathrm{Adj}}({\bf{M}}')$, which links the
stationary vector and the determinant of transition matrix, such
that:
\begin{eqnarray}\label{detandep}
\begin{array}{l}
{\bf{v}}^{T} \cdot {\bf{u}} = \det({{\bf{p}}^1},\cdots,
{{\bf{p}}^X} ,\cdots,{{\bf{p}}^N}, {\bf{u}} ),
\end{array}
\end{eqnarray}
where $({{\bf{p}}^1},\cdots,
{{\bf{p}}^X} ,\cdots,{{\bf{p}}^N}, {\bf{u}})$ is a $ 2^N  \times 2^N $ determinant and ${\bf{u}}$ is
the last column of ${\bf{M}}'$. This theorem is of much significance
since it allows us to calculate one player's long-term expected
payoff by using the Laplace expansion on the last column of
${\bf{M}}'$. Let ${{\bf{u}}^X}$ denote the payoff vector for the
player $X$. Replacing the last column of ${\bf{M}}'$ by
${{\bf{u}}^X}$, we can calculate player $X$'s long-term expected
payoff as:
\begin{equation}
\begin{split}
E^X= \frac {\det({{\bf{p}}^1} ,\cdots,{{\bf{p}}^X},\cdots,
{{\bf{p}}^N} ,{{\bf{u}}^X} )}
           {\det({{\bf{p}}^1},\cdots,{{\bf{p}}^X},\cdots,{{\bf{p}}^N},{\bf{1}} )},
\end{split}
\end{equation}
where $\bf{1}$ is an all-one vector introduced for normalization.
Each player $X$'s expected payoff depends linearly on its own payoff
vector ${\bf u}^X$. Thus making a linear combination of all the
players' expected payoffs yields the following equation:
\begin{equation}\label{mischiefeq}
\begin{split}
&\sum\nolimits_{X=1}^N \alpha_{X}E^X+\alpha_0 =\\
&\frac{\det({{\bf{p}}^1} ,\cdots,{{\bf{p}}^X} ,\cdots,{{\bf{p}}^N} , {\sum\nolimits_{X=1}^N \alpha_{X}{{{\bf
u}^X}}+\alpha_0\mathbf{1}})}{\det({{\bf{p}}^1} ,\cdots,{{\bf{p}}^X} ,\cdots,{{\bf{p}}^N} , {\bf{1}})},\\
\end{split}
\end{equation}
where $\alpha_0$ and $\alpha_{X}$ $(X=1,2,\cdots,N)$ are constants.


This important equation reveals the possible
linear relationship between the players' expected payoffs. Recalling
that in the matrix ${\bf{M}}'$ there exists a column
${\bf{\tilde{p}}}^X$ totally determined by ${{\bf{p}}^X} $, if
player $X$ sets ${{\bf{p}}^X} $ properly and makes
${\bf{\tilde{p}}}^X$ being equal to a linear combination of all the
players' payoff vectors such that:
\begin{equation}\label{eq:Xstrategy}
{\bf{\tilde{p}}}^X=\sum\nolimits_{X=1}^N \alpha_{X}{{{\bf
u}^X}}+\alpha_0\mathbf{1},
\end{equation}
then he can unilaterally make the determinant in Eq.
(\ref{mischiefeq}) vanished and, consequently, enforce a linear
relationship between each player's expected payoff, as:
\begin{equation}\label{eq:linearrelationship}
\sum\nolimits_{X=1}^N \alpha_{X}E^X+\alpha_0 =0.
\end{equation}
Since the determinant of ${\bf{M}}'$ is zero, the strategy
${\bf{p}}^X$ which leads to the above linear equation Eq.
(\ref{eq:linearrelationship}) is a multi-player zero-determinant
strategy of player $X$. Without loss of generality, we assume that
player $1$ is the player adopting ZD strategies, and investigate
the relationship between $1$'s strategy ${\bf{p}}^1$ and its
opponents' total expected payoff $\sum\nolimits_{X = 2}^N {E^X } $.
Hereinafter, the superscript $1$ of $p_{C,n}^1$ and $p_{D,n}^1$ are
all omitted for simplicity.


\section{Iterated Public Goods Games}
Public goods games have been widely studied to examine the behaviors
in the context of social dilemma. In this section, we use the
iterated public goods game as a common paradigm to study the
multi-player ZD strategies. In the public goods games, there are $N$
players who obtain an initial endowment of $c>0$. Without loss of
generality, we set $c=1$. Each player chooses either to cooperate by
contributing the endowment $c=1$ into a public pool, or to defect by
contributing nothing. The total contribution will be multiplied by a
factor $r$ ($1<r<N$) and divided equally among the $N$ players. An
arbitrary player $X$'s payoff at state $i$ then reads
\begin{equation}\label{eq:pggpayoff}
u^X_i = \frac{{r(n(i) + h^X )}}{N} + (1 - h^X),
\end{equation}
where $n(i)$ is the number of cooperators among $X$'s $N-1$
opponents in the state $i$, and $h^X=1$ if player $X$ chooses to
cooperate and $h^X=0$ otherwise. Hence the payoff vector of player
$X$ is ${\bf{u}}^X = [u_1^X,\cdots,u_i^X,\cdots, u_{2^N }^X ]^{T}$.
Figure \ref{ThreePlayer}(b) gives an example of the payoff vectors
for three-player game. We will investigate two kinds of
specializations of ZD strategies, namely pinning strategies and
extortion strategies.

\subsection{Pinning Strategies}
In this paper, when talking about pinning strategies, we mean a
specialization of ZD strategies that can be adopted by a player to
control the total expected payoff of all other $N-1$ opponents,
instead of the expected payoffs of some certain opponents. This is because
as we have mentioned above, in the public goods game, the information
about how many opponents will cooperated is very important while
whether a specific opponent will cooperate is less meaningful. If
the player $1$ wishes to exert a unilateral control over his
opponents' total expected payoff, he can set ${\bf p}^1$ properly
and make ${\bf{\tilde p}}^1$ identical to the last column in the
determinant such that
\begin{equation} \label{eq:zd}
{\bf{\tilde p}}^1  = \mu \sum\nolimits_{X=2}^N {{\bf{u}}^X }  + \xi
{\bf{1}}.
\end{equation}
Then, the determinant will be zero, and a linear function of all
opponents' expected payoffs will be established as:
\begin{equation} \label{eq:E}
\mu \sum \nolimits_{X = 2}^N {E^X } + \xi = 0.
\end{equation}

Note that Eq. (\ref{eq:zd}) consists of a set of $2^N$ equations.
After eliminating the redundancy ones, there remains $2N$
independent linear equations which exactly correspond to the $2N$
independent elements in the strategy vector:
\begin{subequations}  \label{eq:equal_prob}
\begin{align}
 p_{C,n}&=1+\mu \frac{r(n+1)(N-1)+(N-1-n)N}{N}+\xi,           \label{eq:equal_probA} \\
 p_{D,n}&=\mu \frac{rn(N-1)+(N-1-n)N}{N}+\xi, \label{eq:equal_probB}
\end{align}
\end{subequations}
with $n\in\{0, 1, \cdots, {\small{N-1}}\}$. In Eqs.
(\ref{eq:equal_prob}), there are $2N$ probabilities $p_{C,n}$ and
$p_{D,n}$, and the coefficients $\mu$ and $\xi$ are controlled by
player $X$. One can represent all the other $2N-2$ probabilities by
means of $p_{C,N-1}$ and $p_{D,0}$, which are the probabilities for
mutual cooperation and mutual defection, respectively. While
$p_{C,N-1}$ and $p_{D,0}$ themselves are given by:
\begin{subequations} \label{eq:equal_p}
\begin{align}
 p_{C,N-1}&=1+\mu (N-1)r+\xi, \label{eq:equal_pc} \\
 p_{D,0}&=\mu (N-1)+\xi. \label{eq:equal_pd}
\end{align}
\end{subequations}
The parameters $\mu$ and $\xi$ should satisfy the probability
constrains $p_{C,N-1}\in[0, 1]$ and $p_{D,0}\in[0, 1]$. From the two
equations above we can get the allowed value ranges of $\mu $ and
$\xi$. Denote $\mu $ and $\xi$ as follows:
\begin{subequations}\label{eq:equal_muxi}
\begin{align}
\mu&=-\frac{1-p_{C,N-1}+p_{D,0}}{(N-1)(r-1)}, \label{eq:equal_alpha1}\\
\xi&=\frac{1-p_{C,N-1}+rp_{D,0}}{r-1}.\label{eq:equal_alpha0}
\end{align}
\end{subequations}
Introducing $\mu $ and $\xi$ back into Eqs. (\ref{eq:equal_prob}), we
can investigate the feasible regions for all the probabilities
$p_{C,n}$ and $p_{D,n}$. If the probability constrains for all
$p_{C,n}$ and $p_{D,n}$ can be satisfied within $n\in\{0, 1, \cdots,
{\small{N-1}}\}$, it means the pinning strategies exist.

Furthermore, we can also investigate the total expected payoff of
all opponents. Substituting Eqs. (\ref{eq:equal_muxi}) into Eq.
(\ref{eq:E}) yields:
\begin{equation} \label{eq:E2}
\sum\nolimits_{x = 2}^N {E^x }  =  - \frac{\xi }{\mu } = (N - 1) +
\frac{{(r - 1)(N - 1)p_{D,0} }}{{1 - p_{C,N - 1}  + p_{D,0} }}.
\end{equation}
Hence, the opponents' total expected payoff is still determined only
by $p_{C,N-1}$ and $p_{D,0}$. If $p_{C,N-1}$ and $p_{D,0}$ satisfy a
linear relationship $\gamma p_{D,0}  + p_{C,N - 1}  - 1 = 0$ (i.e.,
$\gamma = \frac{{1 - p_{C,N - 1} }}{{p_{D,0} }} $), then Eq.
(\ref{eq:E2}) can be rewritten as:
\begin{equation} \label{eq:E3}
\sum\nolimits_{x = 2}^N {E^x }  = (N - 1) + \frac{{(r -
1)(N - 1)}}{{1 + \gamma }}.
\end{equation}
The opponents' total expected payoff then depends only on the number
of players $N$, the multiplication factor $r$, and the parameter
$\gamma$.

After combination and reduction, Eq. (\ref{eq:equal_probA}) and Eq.
(\ref{eq:equal_probB}) can be written in the following format:
\begin{subequations}  \label{eq:equal_prob_reform}
\begin{align}
 p_{C,n}  = 1 + \frac{\mu }{N}\left\{ {\left[ {r\left( {N - 1}
\right) - N} \right]n + \left( {N - 1} \right)\left( {r + N}
\right)} \right\} + \xi,           \label{eq:equal_probA_reform} \\
 p_{D,n}  = \frac{\mu }{N}\left\{ {\left[ {r\left( {N - 1} \right) -
N} \right]n + \left( {N - 1} \right)N} \right\} + \xi ,
\label{eq:equal_probB_reform}
\end{align}
\end{subequations}
in which $N$, $r$ are constance if the game setting is fixed. We can
see in the above two inequations, a comment term referring to
variable $n$ is $ \mu n \left(r - \frac{N}{{N - 1}}\right)\left( {N - 1}
\right) $. $p_{C,n}$ and $p_{D,n}$ are functions with variable
$n$, and their monotonicity is determined by $n$'s coefficients
$\mu$ and $ \left(r - \frac{N}{{N - 1}}\right)$. So let us discuss
about different cases of $ \left(r - \frac{N}{{N - 1}}\right)$.

\textbf{\emph{Case 1}.} When $r<\frac{N}{N-1}$, Eqs.
(\ref{eq:equal_prob}) are monotonously increasing functions of $n$,
It is then sufficient to check $p_{C,n}$ and $p_{D,n}$ at the lower
bound and upper bound of $n$. Since $p_{C,N-1}$ and $p_{D,0}$ should be selected
in the feasible region, we need only to check $n=0$ for $p_{C,n}$
and {\small{$n=N-1$}} for $p_{D,n}$. Then the probability constrains become:
\begin{subequations}\label{eq:equal_prob2}
\begin{align}
p_{C,0}&=1+\mu (N-1+r\frac{N-1}{N})+\xi\geq0, \label{eq:equal_prob22} \\
p_{D,N-1}&=\mu r\frac{(N-1)^2}{N}+\xi\leq1. \label{eq:equal_prob23}
\end{align}
\end{subequations}
By substituting Eqs. (\ref{eq:equal_muxi}) into Ineqs.
(\ref{eq:equal_prob2}), we have
\begin{subequations}\label{eq:probConstrainEq3}
\begin{align}
rp_{C,N-1}+(rN-N-r)p_{D,0}-r+rN-N &\geq0, \label{eq:probConstrainEq31}\\
(rN-N-r)p_{C,N-1}+rp_{D,0}-2rN+r+2N &\leq0.\label{eq:probConstrainEq32}
\end{align}
\end{subequations}
The two feasible half-planes respectively constituted by Ineq.
(\ref{eq:probConstrainEq31}) and Ineq. (\ref{eq:probConstrainEq32})
intersect at the point
\begin{equation} \label{case1-intersection}
(p_{C,N-1}^*,p_{D,0}^*)=\left(1-\frac{rN-N}{N+2r-rN},\frac{rN-N}{N+2r-rN}\right).
\end{equation}

Obviously, $p_{C,N-1}^*$ and $p_{D,0}^*$ satisfy the linear relationship $p_{C,N-1}^*+p_{D,0}^*=1$, and it is easy to validate that $p_{D,0}^*<0$ when $r<1$, implying that there is no feasible region for pinning strategies for $r<1$.

When $r=1$, the point $(p_{C,N-1}^*,p_{D,0}^*)=(1,0)$, which is the unique feasible point.
From Eq. (\ref{eq:equal_probA}) and Eq. (\ref{eq:equal_probB})
it can be found that $\mu = 0$ and $\xi = 0$ when $(p_{C,N-1},p_{D,0})=(1,0)$,
where the singular strategy is $p_{C,n} = 1$ and $p_{D,n}
= 0$ for $n \in \{ 0,1,...,N - 1\} $. Under such case, to enforce a
pinning strategy, a player should always cooperate once he starts
the game with cooperation, or, always defect once he starts the game
with defection. The expected probability he will take $C$ or $D$ depends on the initial 
probability distribution over his pure strategy space. Then, the state transition
matrix $\bf M$ in Eq. (\ref{eq:transmatrix}) becomes a block
diagonal matrix with two closed communicating classes, which
indicates that the Markov chain's stationary distribution is not unique
(i.e. depending on the initial distribution), suggesting
that this transition matrix does not essentially have a stationary
distribution with respect to a unit eigenvalue. Consequently, in the
case of $r = \frac{N}{{N - 1}}$, the expected payoff cannot given by
the determinant form as proposed by Press and Dyson \cite{Press2012}.

\begin{figure}
\centerline{\includegraphics[width=0.75\linewidth,height=0.75\linewidth]{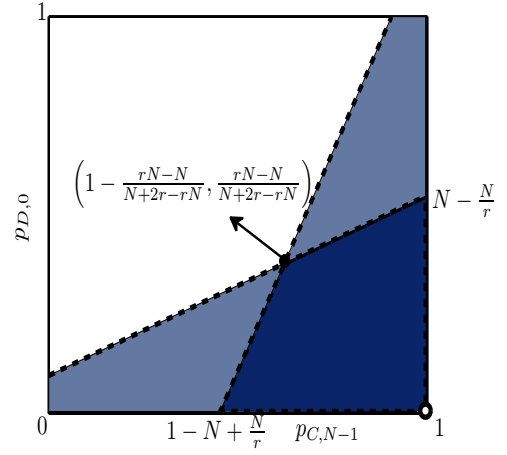}}
\caption{The feasible region of the pinning strategies when $1<r <
\frac{N}{N-1}$, which is determined by the intersection of the two
half-planes formed in terms of the two linear inequalities in
(\ref{eq:probConstrainEq3}), except for the singular point
$(p_{C,N-1},p_{D,0})=(1,0)$. }\label{feasible2}
\end{figure}

When $1<r< \frac{N}{N-1}$, the conditions $0<p_{C,N-1}^*<1$
and $0<p_{D,0}^*<1$ are ensured, which means there always exists a
feasible region for pinning strategies. The corresponding feasible region is emphasized by
dark blue, as shown in Fig.~\ref{feasible2}. Then, the minimum value of all opponents' total expected payoff can be reached when $p_{D,0}=0$ and $p_{C,N
- 1} \ne 1$:
\begin{equation}
\left( {\sum\nolimits_{x = 2}^N {E^x } } \right)_{\min }  = (N - 1).
\end{equation}
If $p_{C,N-1}=1$ and $p_{D,0} \ne 0$, the maximum value is:
\begin{equation}
\left( {\sum\nolimits_{x = 2}^N {E^x } } \right)_{\max }  = r(N -
1).
\end{equation}
Therefore, the player $1$ can pin his opponents' average expected
payoff to the range between $1$ and $r$ when $r < \frac{N}{{N -
1}}$.

\begin{figure}[h]
\centerline{\includegraphics[width=0.75\linewidth,height=0.75\linewidth]{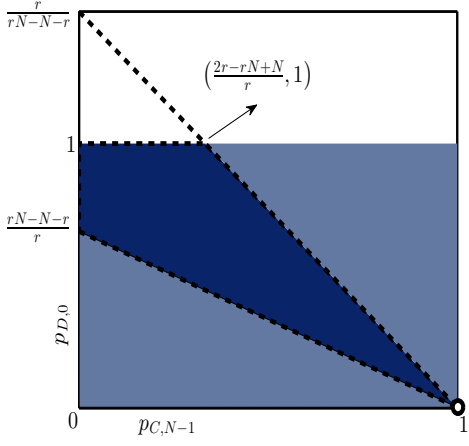}}
\caption{The feasible region of the pinning strategies when
$r>\frac{N}{N-1}$, which is determined by the intersection of the
two half-planes formed by the two linear inequalities in
(\ref{eq:probConstrainEq2}). The intersected region is a convex hull
with four extreme points. The region shrinks while the gradients of
the two confine lines approaches each other.}\label{feasible}
\end{figure}

\textbf{\emph{Case 2}.} When $r = \frac{N}{{N - 1}}$, the intersecting point reads
$(p_{C,N-1}^*,p_{D,0}^*)=(0,1)$ and a pinning strategy can be obtained
through arbitrarily selecting $p_{C,N-1}$ and $p_{D,0}$
in the region of $[0,1]$ except for the singular point $(p_{C,N-1},p_{D,0})=(1,0)$. Along the line $\gamma p_{D,0}  + p_{C,N - 1} = 1$, 
the opponents' total excepted payoff can be pinned into the value determined by Eq. (\ref{eq:E3}), dependent on the parameters $N$, $r$ and $\gamma$. The maximum and minimum values of player $1$'s excepted payoff occurs when all opponents choose always-C and always-D strategies, respectively.

\textbf{\emph{Case 3}.} When $r>\frac{N}{N-1}$, Eqs.
(\ref{eq:equal_prob}) are monotonously decreasing functions of $n$.
It is thus sufficient to check the maximum value $p_{C,0}$ and the minimum value $p_{D,N-1}$.
Then the probability constrains becomes:
\begin{subequations}\label{eq:probConstrainEq1}
\begin{align}
p_{C,0}&=1+\mu (N-1+r\frac{N-1}{N})+\xi\leq1, \label{eq:probConstrainEq12}\\
p_{D,N-1}&=\mu r\frac{(N-1)^2}{N}+\xi\geq0.
\label{eq:probConstrainEq13}
\end{align}
\end{subequations}
Following a similar procedure as Case 1,
by substituting Eqs. (\ref{eq:equal_muxi}) into Eqs.
(\ref{eq:probConstrainEq1}), we can get:
\begin{subequations}\label{eq:probConstrainEq2}
\begin{align}
rp_{C,N-1}+(rN-N-r)p_{D,0}-r &\leq0, \label{eq:probConstrainEq21}\\
(rN-N-r)p_{C,N-1}+rp_{D,0}-rN+r+N &\geq0,
\label{eq:probConstrainEq22}
\end{align}
\end{subequations}
each of which constitutes a closed half-plane in the two-dimensional
real space $\mathbb{R}^2$. These two half-planes intersect at the dark blue
region in Fig.~\ref{feasible}, with four extreme points $ (0,1) $, $
(0,\frac{{rN - N - r}}{r}) $, $(\frac{{2r - rN + N}}{{r}},1)$ and $
(1,0)$. The feasible region converges to a line $p_{C,N-1}+p_{D,0}=1$
when $\frac{rN-N-r}{r}=\frac{r}{rN-N-r}$, i.e., $r=\frac{N}{N-2}$.
The feasible region for the pinning strategies vanishes when $r>\frac{N}{N-2}$.
Meanwhile, considering $r>\frac{N}{N-1}$, now we
have the two boundaries, as
\begin{equation}
\frac{N}{N-1}<r\le\frac{N}{N-2}.
\label{boundarycase2}
\end{equation}

According to Eq. (\ref{eq:E3}), we can obtain the minimum and
maximum values of the opponents' total expected payoff in the case of $r
> \frac{N}{{N - 1}}$:
\begin{subequations}
\begin{align}
\left( {\sum\nolimits_{x = 2}^N {E^x } } \right)_{\min } \left|
{_{\gamma  = \frac{r}{{rN - N - r}}} } \right. = r(N - 2 +
\frac{1}{N}),
\\
\left( {\sum\nolimits_{x = 2}^N {E^x } } \right)_{\max } \left|
{_{\gamma  = \frac{{rN - N - r}}{r}} } \right. = (N - 1) + r(1 -
\frac{1}{N}).
\end{align}
\end{subequations}
When $r=\frac{N}{N-2}$, $\sum\nolimits_{x = 2}^N {E^x }  = \frac{(r+1)^2}{2(r-1)}.$

\begin{figure}[h]
\centerline{\includegraphics[width=0.75\linewidth,height=0.75\linewidth]{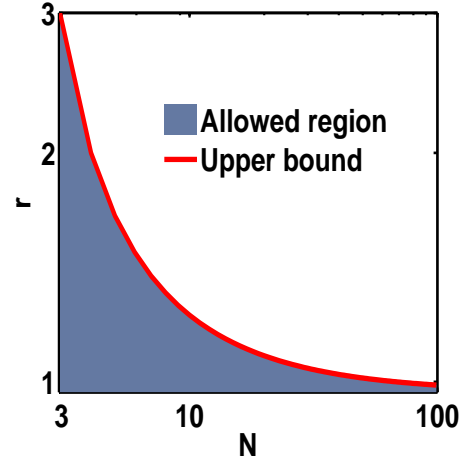}}
\caption{Log-log plot of the upper bound of $r$. The upper bound is
a monotonously decreasing function of the group size $N$, namely
with the increasing of $N$, the allowed region of multiplication
factor for a pinning strategy shrinks.}\label{upperboundeq}
\end{figure}

In summary, given the multiplication factor $r$, if player $1$ wants to pin the total
expected payoff of all other opponents, it is required that
$N\le\frac{2r}{r-1}$. Or, in a fixed group size $N$, the player $1$
can do this only when $1<r\le\frac{N}{N-2}$.
The upper bound of $r$ as a function of the group
size $N$ is presented in Fig.~\ref{upperboundeq}. Pinning the total
expected payoff of all other opponents is becoming difficult as $N$
grows. Figure 5(a) shows an example 3-player IPGG, where the ZD player $X$ can pin his opponents' total expected payoff into a fixed value, while his own payoff depends on the opponents' behaviors: if he would like to set a high value, he may lose more.

In the above analysis, one player's strategy is only conditioned on how
many of his opponents cooperate while the detailed information about who
cooperate is less important. Such settings may essentially reduce
the constrains for the existence of pinning strategies. When
considering more complicated scenarios, the constrains necessarily
become more strict. For example, there may be more linear
inequalities to be satisfied in Eqs. (\ref{eq:probConstrainEq2}) and Eqs.
(\ref{eq:probConstrainEq3}). Each inequality constitutes a
half-space in the $m$-dimensional real space $\mathbb{R}^m$ where $m$ is
the number of probability variables in the inequality set. Finding
the feasible region of pining strategies is then transferred to the calculation of the
intersections of these $m$ half-spaces, which is equivalent to a traditional linear
programming problem. Furthermore, since the feasible region for a
pining strategy is essentially a convex hull, when analyzing the
properties of the pinning strategies, it is sufficient to concentrate
on the extreme points. Such feature brings us convenience to further study the game's
equilibriums.

\begin{figure*} 
\subfigure[ ] { \label{fig5:a}
\includegraphics[width=0.23\linewidth,height=0.24\linewidth]{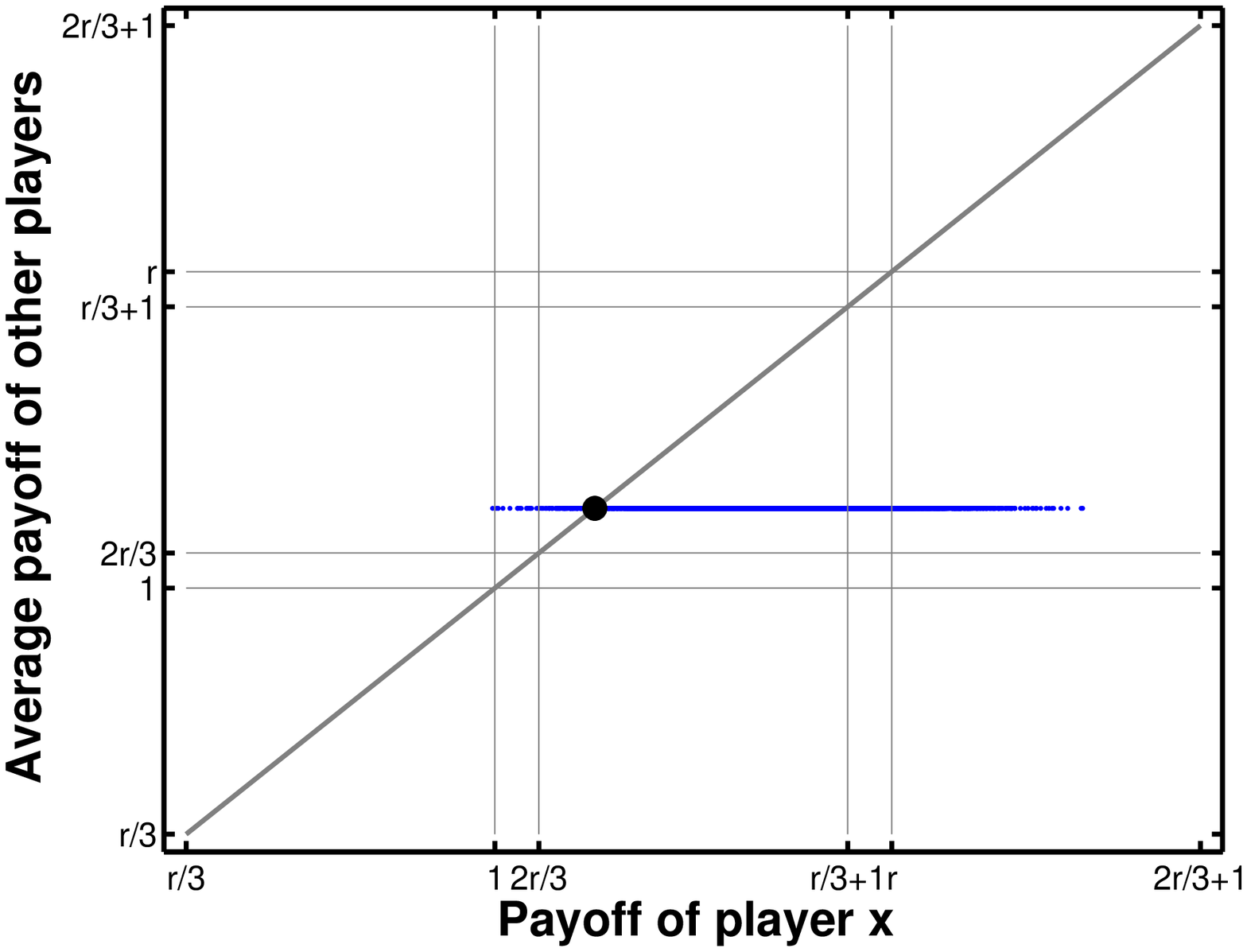}
} \subfigure[ ] { \label{fig5:b}
\includegraphics[width=0.23\linewidth,height=0.24\linewidth]{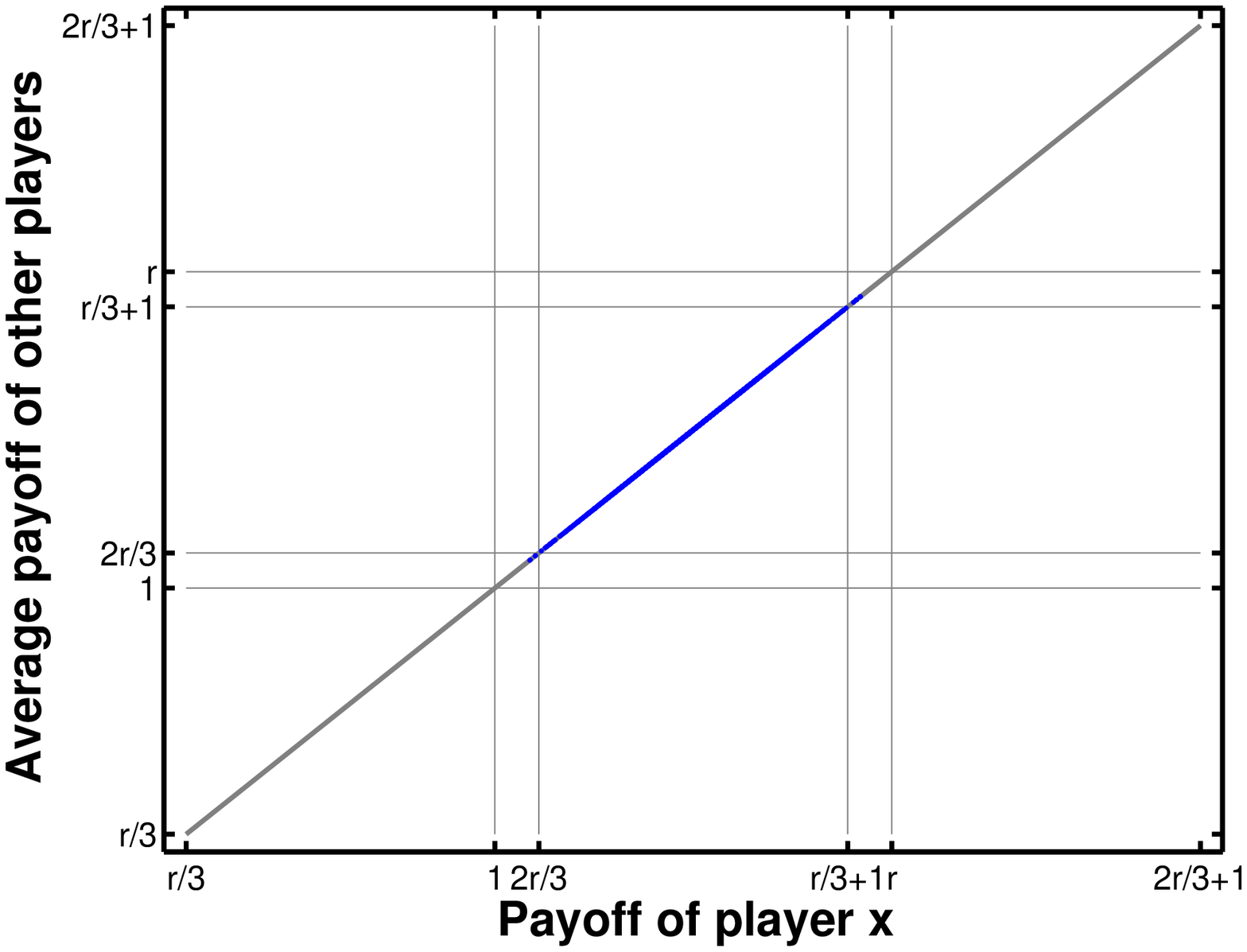}
} \subfigure[ ] { \label{fig5:c}
\includegraphics[width=0.23\linewidth,height=0.24\linewidth]{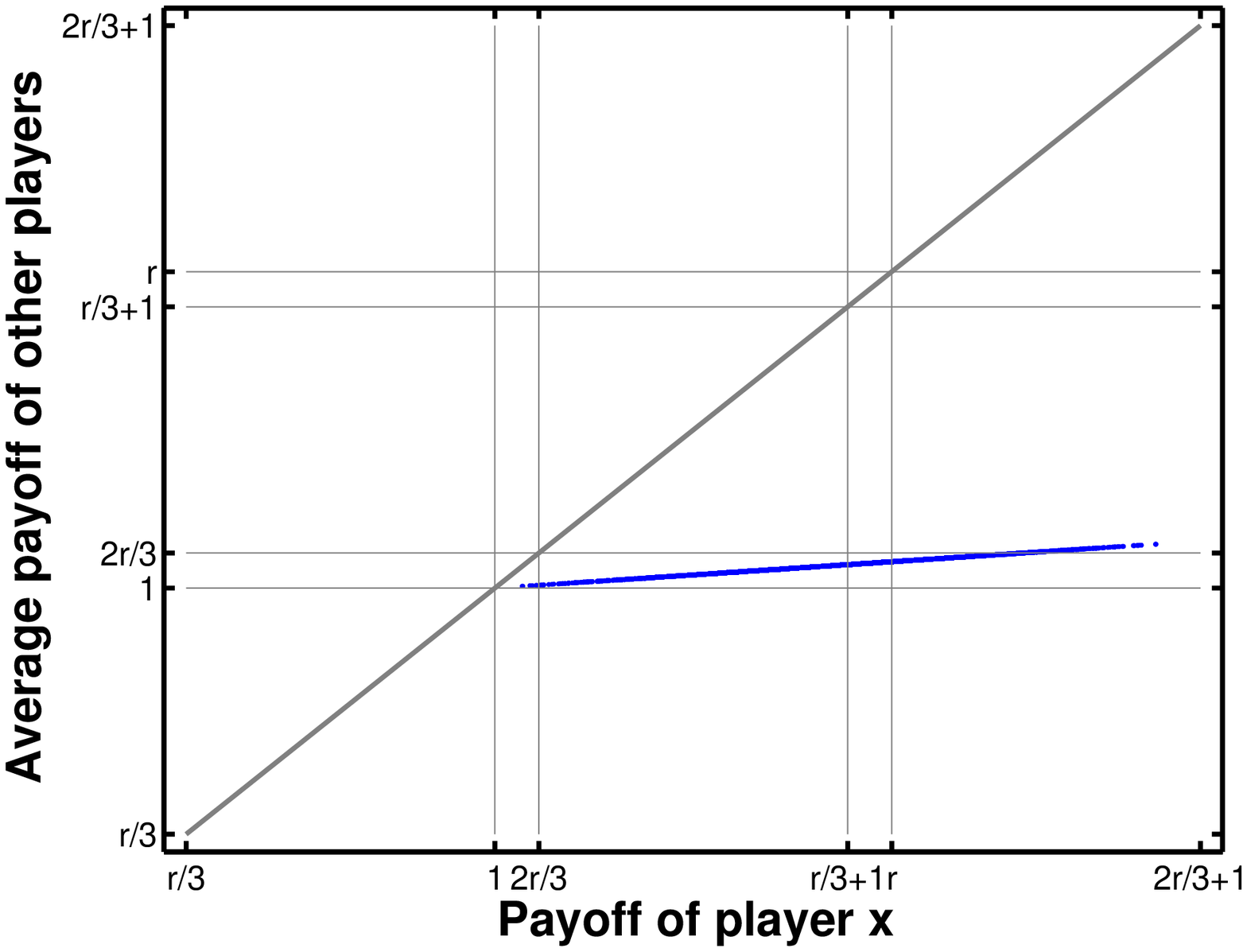}
} \subfigure[ ] { \label{fig5:d}
\includegraphics[width=0.23\linewidth,height=0.24\linewidth]{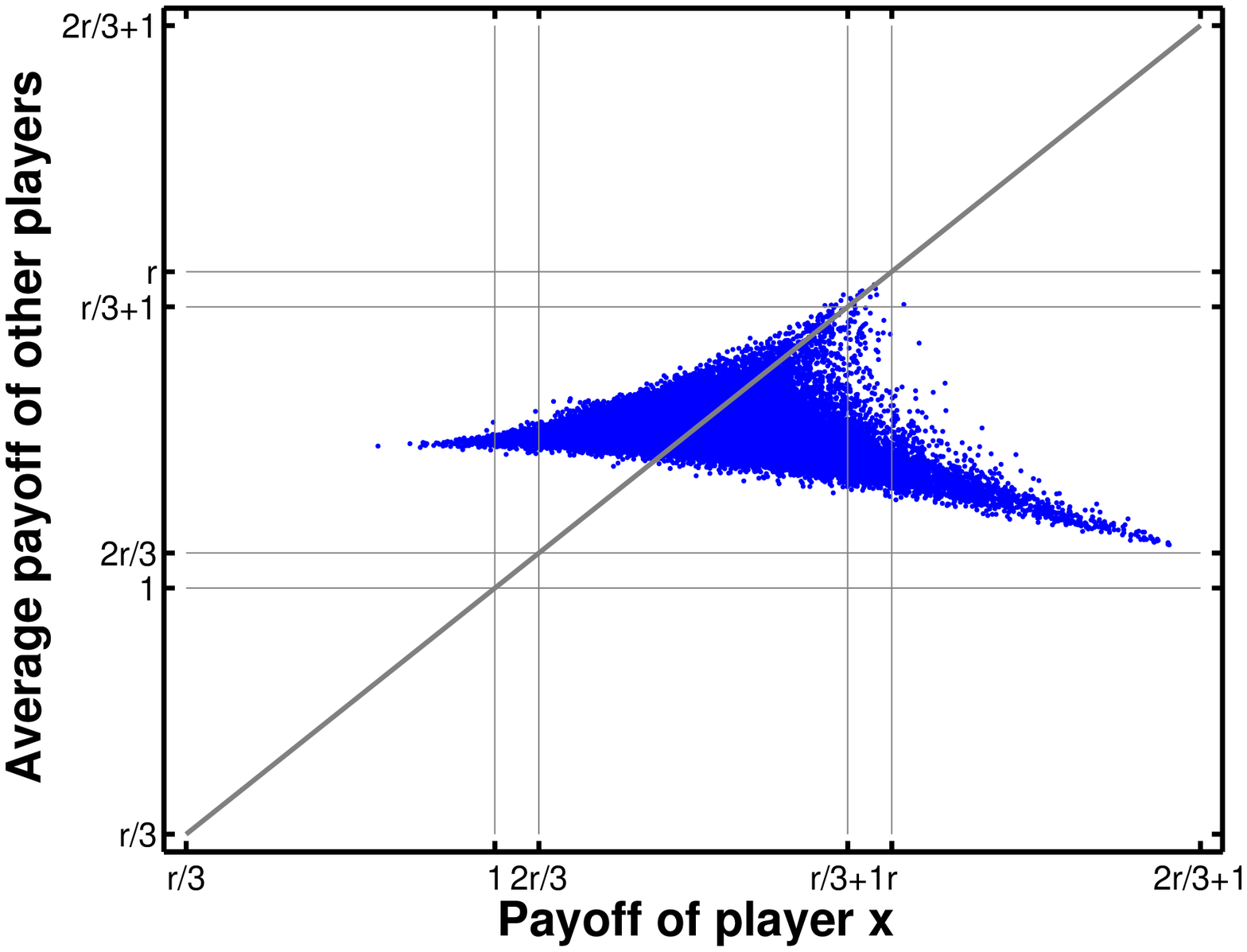}
}\caption{The payoff of the focal player $X$ against the average
payoff of the other two players in an example $3$-player IPGG with $r=1.6$ fixed. (a)
The pinning strategy $ {\bf{p}}^X=[0.08, 0.15, 0.15, 0.22, 0.17,
0.24, 0.24, 0.31]^{T}$, corresponding to the state sequence of $CCC$,
$CCD$, $CDC$, $CDD$, $DCC$, $DCD$, $DDC$ and $DDD$. (b) The
$\chi$-extortion strategy with the minimal $\chi=0.5$ (i.e., $\chi(N-1)=1$) and
${\bf{p}}^X=[1.0, 0.9, 0.9, 0.8, 0.2, 0.1, 0.1, 0.0]^{T}$, where the ZD player $X$ fairly shares the surplus with his opponents. (c) The
$\chi$-extortion strategy with $\chi=7.9$ and ${\bf{p}}^X=[0.87, 0.87,
0.87, 0.86, 0.01, 0, 0, 0]^{T}$, which is close to the upper bound
$\chi=8$. (d) The Win-Stay-Lose-Shift (WSLS) strategy with ${\bf{p}}^X=[1, 0,
0, 0, 0, 1, 1, 1]^{T}$ for comparison, where the average payoff of the two opponents will distribute a two-dimensional area. Each plot reports the aggregated results from $10^6$ independent stage games, where each data point represents the result of one stage game. In each stage game, the strategies of $X$'s two opponents are assigned randomly.} \label{FigStrategy}
\end{figure*}

Our analysis indicates that, in an $N$-player IPGG with proper
settings, one player can unilaterally control the opponents' total expected
payoff and pin it to a fixed value by playing ZD strategies. In this
case, the pinned total expected payoff of opponents is no less than
the total endowment of them, which indicates that the player using
pinning strategy is nice and may run risk to decrease his own payoff (as
indicated by the short horizontal line to the left of the diagonal line in Fig. \ref{FigStrategy}(a)). Generally, in an IPGG, a player
can pin the opponents' total expected payoff when the group size $N$
does not exceed an upper bound, or, when the multiplication factor
is not too large. That is to say, the condition for pinning multiple players'
total expected payoff in the IPGG is more strict than pinning a
single opponent's expected payoff in a two-player IPD. In Appendix A, we prove that, in the multi-player IPGG, a ZD
player cannot unilaterally set his own expected payoff, analogous to the two-player IPD. In Appendix B, we show that in the multi-player IPGG, two or more players cannot collusively control other players' payoff.

\subsection{Extortion Strategies}
Besides pinning the opponents' total payoff, a ZD player can also
extort all his opponents and guarantee that his own surplus over the
free-rider's payoff is $\chi$-fold of the sum of opponents' surplus.
This is the so-called $\chi$-extortion strategy, where $\chi$ is the
extortionate ratio. Formally, the extortion strategies for a ZD player
$1$ is:
\begin{subequations}\label{eq:extortionate}
\begin{align}
\tilde{{\bf{p}}}^1=\Phi\left[({\bf{u}}^1-\mathbf{1})-\chi
\sum_{X=2}^N ({\bf{u}}^X-\mathbf{1})\right ].
\end{align}
\end{subequations}
Solving this vector equation gives us $2N$ linear equations:
\begin{subequations}\label{eq:extortionate1}
\begin{align}
p_{C,n}  = 1 + \Phi \left( {\frac{{rn}}{N} - \chi \frac{{
rn(N - 1)- nN}}{N}} \right) \notag \\
+ \Phi \left( {\frac{{r - N}}{N} - \chi \frac{{r(N - 1)}}{N}}
\right),\label{eq:extortionate11} \\
p_{D,n}  = \Phi \left( {\frac{{rn}}{N} - \chi \frac{{ rn(N - 1)-
nN}}{N}} \right), \label{eq:extortionate12}
\end{align}
\end{subequations}
for {\small{$n\in \{ 0,1,\cdots,N-1 \} $}} and $p_{C,n} ,p_{D,n}  \in [0,1]$. 

When $ \chi  \ge 0 $, $ {\frac{{r - N}}{N} - \chi
\frac{{r(N - 1)}}{N}} $ is always negative. If $\Phi < 0$, the term
in the bracket in Eq. (\ref{eq:extortionate12}) should be
non-positive to make $p_{D,n} \ge 0$, leading to $p_{C,n} >
1$, which is out of the probability range. The case of $\Phi = 0$
corresponds to the singular strategy of $p_{C,n} = 1$ and $p_{D,n}
= 0$ for $n \in \{ 0,1,...,N - 1\} $. Thus it is required that
$\Phi > 0$. In this case, we have the following linear
inequalities as constrains for the extortion strategies:
\begin{subequations}\label{eq:extortionateineqs}
\begin{align}
\frac{r(n+1)-N}{N}-\chi\frac{r(n+1)(N-1)-nN}{N}\leq0, \label{eq:extortionateineqs1}\\
\frac{rn}{N}-\chi\frac{rn(N-1)-nN}{N}\geq0.
\label{eq:extortionateineqs2}
\end{align}
\end{subequations}
Given $n$, $\chi$ is determined by both $r$ and $N$. This is
different from the two-player IPD \cite{Press2012} where $\chi$ can
take any value. From the above two sets of constrains for $\chi$, if $r \le \frac{N}{N-1}$,
\begin{equation}\label{eq:extortionate6}
\chi \ge \frac{1}{N-1}.
\end{equation}
else, if $r>\frac{N}{N-1}$,
\begin{equation}\label{eq:extortionate5}
\frac{1}{N-1}\leq\chi\leq \frac{r}{r(N-1)-N}.
\end{equation}
Figures \ref{FigStrategy}(b) and \ref{FigStrategy}(c) show the numerical example of extortion strategies. Within the allowed range of $\chi$ (as shown in the above inequalities), the average payoff of all other opponents falls in a line.

\begin{figure}[h]
\centerline{\includegraphics[width=1.1\linewidth,height=0.65\linewidth]{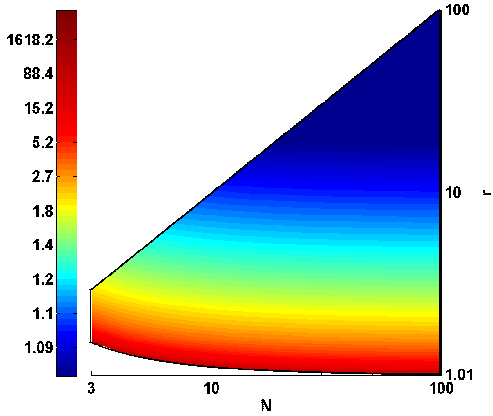}}
\caption{The upper bounds of $\chi(N-1)$ (represented by different colors) versus $N$ 
 and $r$, in the region $r>\frac{N}{N-1}$ and $N\geq 3$. 
}\label{boundex}
\end{figure}

For any value of $r$, $\chi$ has its lower bound. When
$r>\frac{N}{N-1}$, $\chi$ also has its upper bound $\frac{r}{r(N-1)-N}$. Normalizing $\sum_{X=2}^N ({\bf{u}}^X-{\bf {1}})$ by the number of opponents $(N-1)$, the ZD player can extort over the average payoff of his
opponents by an effective ratio $\chi(N-1)$, which has an upper bound
$\frac{r(N-1)}{r(N-1)-N}$. For sufficiently large $N$,
\begin{equation} \label{chimax}
\mathop{\lim}_{N\rightarrow\infty}\chi_{\max}(N-1)=\frac{r}{r-1}.
\end{equation}
In Fig.~\ref{boundex}, we show the value of $\chi_{\max}(N-1)$ as a function of the group size $N$ and the multiplication factor $r$, in the region $r>\frac{N}{N-1}$. As shown in Eq.~(\ref{chimax}) and Fig.~\ref{boundex}, for large $N$, $r$ can be very close to 1, leading to a large maximum effective extortionate ratio $\chi_{\max}(N-1)$. However, under such case, a player is usually not willing to cooperate and thus the payoff in addition to the endowment is tiny. That is to say, although the effective extortionate ratio can be huge, the extorted payoff is not much.

Substituting the bounds of $\chi$ into the probabilistic
strategies in Eqs. (\ref{eq:extortionate1}), we can obtain the
allowed range for $\Phi$ as:
\begin{subequations}\label{eq:phi-bound}
\begin{align}
\Phi\leq\frac{1}{\chi\frac{r(n+1)(N-1)-nN}{N}-\frac{r(n+1)-N}{N}}, \label{eq:phi-bound1}\\
0 < \Phi \leq\frac{1}{\frac{rn}{N}-\chi\frac{rn(N-1)-nN}{N}}.
\label{eq:phi-bound2}
\end{align}
\end{subequations}
According to the monotonicity, these two inequalities can be reduced
to:
\begin{equation}
\begin{split}
0< \Phi \leq \frac{N}{N-r+\chi r(N-1)}.
\end{split}
\end{equation}

Note that $\frac {N}{N-1}$ is monotonously decreasing with $N$. Thus
given a specific multiplication factor $r$, the extortionate ratio
$\chi$ is more likely to have an upper bound when more players are
involved in the game. This means in a game with more players, it
will be more difficult for the extortioner to secure his own payoff
by using ZD strategy and setting a fixed ratio between his and the
opponents' surplus. A tricky strategy of the extortionate player thus
will be restrained when he plays with more opponents. On the other hand, given a
fixed group size, a large $r$ will shrink the feasible range of the
extortionate ratio. A large multiplication factor $r$ results in a
better reward for each player, which promotes mutual cooperations.
Therefore, the above analysis reveals the significant fact that, to
reduce the possible injuries from a crafty egoist, increasing the
cooperation incentive $r$ is an effective approach.

\section{Conclusion and Discussions}

The discovery of ZD strategies makes us both excited and worried, since a selfish person seems to have a more powerful mathematical tool to extort payoffs from those kindhearted and simpleminded people. Although some recent works \cite{Adami2013,Hilbe2013,Stewart2013} suggested that the extortion strategies in two-player IPD are not evolutionary stable, a few ZD players can still extort other non-ZD opponents in a population. Indeed, compared with those well-known game strategies \cite{Axelrod1984}, the ZD strategies are too complicated to be mastered by normal persons, who will eventually become exploitees in the present of ZD players.

To explore the general applicability and limitations of ZD strategies, we have taken a step from two-player games to multi-player games, with the iterated public goods game being the selected template. The bad news learned from our study is that a single ZD player can unilaterally pin the total expected payoff of all other opponents and extort them by enforcing a linear relationship between his own payoff and the opponents' total payoff. A good news from the results is that the capacity of a ZD player to either pin or extort other opponents is more strictly limited compared with the two-player games. Roughly speaking, we can suppress the influences of the ZD player by increasing the number of participants and/or encouraging cooperation via enlarging the multiplication factor. Taking the global warming problem as an example, if we have made more people being aware of the seriousness of such issue and understanding that the abandonment of some environmentally costly lifestyles is of great significance for the sustainable development, we can to some extent enlarge $N$ and $r$ and thus suppress ZD players. Another good point is that when there are more than one ZD players in the IPGG, they cannot collusively control others but each fights his own battle.

Iterated games with private monitoring represent long-term
relationships among players where each player privately receives a
noisy (imperfect) observation of the opponents' actions
\cite{Nowak1993}. The difficulty of handling such games comes from
the fact that players do not share common information under private
monitoring, and the decision making in such games involves with
complicated statistic inference. Consequently, the analysis,
optimization, cooperation enforcement and control in such games have
been known as long-standing challenges. This subclass of game theory has
found a wide range of applications \cite{Mailath2006}, such as evolution in
a realistic noisy environment \cite{Hansen2004,MowakEvo2006}
and agent planning under uncertainty \cite{Phelan2012}. Whether ZD strategies still
works in the noisy environments? Is it still possible for a crafty
egoist to control the payoffs of his opponents? These questions ask for future in-depth understanding of ZD strategies.

Both the origin of life and the formation of human societies require
cooperation \cite{Apicella2012,VaidyaNature2012}. During the
history of biological evolution, animals and microorganisms such as
vampire bats, three-spined sticklebacks, cleaner fishes and bacteria
can recognize the importance of reciprocity and even cooperate
according to tit-for-tat strategy \cite{WilkinsonNature1984,
MilinskiNature1987, BsharyNature2008, LeeNature2010}. Male
side-blotched lizard and \emph{ Escherichia coli} can play the
rock-paper-scissors game in order to maintain biodiversity
\cite{SinervoNature1996, KerrNature2002}. Human is the champion of
cooperation. With the growth, children may change from selfishness
to egalitarian \cite{FehrNature2008}. It is worth exploring whether we can find some field evidences that human beings and animals may be already aware of the
existence of ZD strategies during the biological evolution. 

Researchers can also design laboratory experiments and study
responses of human beings when facing ZD strategies \cite{Rand2013}. A
player may vary his strategy frequently that cannot generate a
Markovian stationary state. Therefore, there are some interesting
problems such as whether some proper ZD strategies can control
opponents' payoff in a short timescale and how a smart player alters
his ZD strategies in terms of his opponents' responds. And of course, we firstly want to know whether a normal person will become crazy when facing a crafty ZD player.

Furthermore, for a large population, an individual cannot interact
with everybody else. Some individuals usually interact more often
than others. The spatial structures of population may affect the
maintenance of cooperation. Then some questions natural arise, for
example, what is the relationship between the different population
structures and related ZD strategies and whether the cooperation can
sustain in dynamic social network with the evolution of ZD
strategies \cite{Santos2011, Hauert2002, SzaboPRL2002, Santos2008,
RongPRE2010, Apicella2012, Perc2013, RandPNAS2011}. Network analysis is then expected to plays a significant role \cite{SzaboFath2007}.

Different from the Prisoner's Dilemma game which characterizes the
pairwise interaction, the public goods game depicts the group
interaction. In the pairwise Prisoner's Dilemma game, only two
players take part in one game. If both of them are extortioners,
their surpluses become zero that leads to the evolutionary
instability of extortion strategies in an infinite population.
However, for the public goods game, $N$ players participate in one
game. It is difficult to ensure all players are aware of the
existence of ZD strategies and use the extortion strategies. Hence,
comparing with the pairwise Prisoner's Dilemma game, the situation for public goods game is more complicated when considering the
evolutionary stability. It is relatively easy to analyze the
evolutionary stability of ZD versus special strategies, such as
always cooperation, always defection, win-stay-lose-shift, and so on. Since the strategy
space of public goods game is very huge comparing with Prisoner's
Dilemma game, we should carefully consider how to perform
Monte Carlo simulations of population in the framework of weak
mutation similar to Ref. \cite{Hilbe2013}. Moreover, in this
paper we only consider the extortion strategies. Recently good
strategies in IPD have been studied \cite{Stewart2013, Akin2012},
which can to be extended to multi-player IPGG by replacing
$\bf{1}$ in Eqs. (\ref{eq:extortionate}) with $\bf{r}$. Then the
robustness of good and generosity strategies can be deeply analyzed in the
next step. In addition, the evolutionary stability analysis of IPGG can also be
combined with the studies on the effects of reward and punishment
\cite{SigmundPNAS01,Sigmund2010,Sasaki2012}.

\begin{acknowledgments}
The authors acknowledge the valuable suggestions and comments from Guan-Rong Chen, Petter Holme,
Gang Yan and Qian Zhao. This work was partially supported by the
National Natural Science Foundation of China (NNSFC) under Grant Nos.
61004098 and 11222543, the Program for New Century Excellent Talents in University under
Grant No. NCET-11-0070, and the Special Project of Sichuan Youth Science and Technology Innovation Research Team under Grant No. 2013TD0006.
\end{acknowledgments}

\begin{appendix}

\section{$X$ tries to set his own payoff}
A ZD player $X$ cannot unilaterally set his own payoff in the PD,
here we obtain the same conclusion for iterated PGG. If he tries to
set his own payoff, he must choose $\tilde{p}^X=\alpha_1
S_1+\alpha_0\mathbf{1}$. The linear equations now become
\begin{equation}\label{eq:equal_prob1S}
p_{C,n}=1+\alpha_1\frac{r(n+1)}{N}+\alpha_{0},
\end{equation}
\begin{equation}\label{eq:equal_prob2S}
p_{D,n}=\alpha_1\frac{rn+N}{N}+\alpha_{0}.
\end{equation}
Setting $p_{C,N-1}$ and $p_{D,0}$ as free variables, we have
\begin{subequations}\label{eq:alphaeS}
\begin{align}
\alpha_1&=\frac{p_{C,N-1}-p_{D,0}-1}{r-1}.\\
\alpha_0&=\frac{rp_{D,0}-p_{C,N-1}+1}{r-1}.
\end{align}
\end{subequations}
Since $p_{C,n},p_{D,n}$ are decreasing functions of $n$, we have
\begin{subequations}\label{eq:probConstrainSelf}
\begin{align}
p_{C,N-2}&=1+\alpha_1 \frac{r(N-1)}{N}+\alpha_0\geq0 \\
p_{C,0}&=1+\alpha_1 \frac{r}{N}+\alpha_0\leq1 \label{eq:probConstrainSelf2}\\
p_{D,N-1}&=\alpha_1 \frac{r(N-1)+N}{N}+\alpha_0\geq0 \\
p_{D,1}&=\alpha_1 \frac{r+N}{N}+\alpha_0\leq1.
\end{align}
\end{subequations}
After some algebra, (\ref{eq:probConstrainSelf2}) can be reduced to
\begin{equation}
p_{D,0}\leq\frac{(N-r) p_{C,N-1}-(N-r)}{rN-r}.
\end{equation}
Since $r\leq N$ for PGG, this leads to $p_{D,0}\leq0$. So a ZD player $X$ cannot set his
own payoff.

\section{Collusive strategies}
In the determinant form of player's payoff, there are columns which
are controlled by more than one players. This suggests that there
might be collusive strategies, which means more than one players
trying to control other players' payoff collusively. However this
type of ZD strategies generally does not exit. Take the $two-$player
collusive strategies as an example. Denote the column controlled
tangly by the player $X$ and $Y$ by $\tilde{pq}$. For general ZD
strategies $\tilde{pq}=\sum_{X=1}^N \alpha_{X} {\bf
S}_X+\alpha_0\mathbf{1}$, the following $2N$ linear equations must
be satisfied: $p_{C,n}q_{C,n}=1+\Theta_1$, $p_{D,n}q_{D,n}=\Theta_2$, $p_{C,n}q_{D,n}=\Theta_3$, and $p_{D,n}q_{C,n}=\Theta_4$. Here $\Theta_1$,$\Theta_2$,$\Theta_3$ and $\Theta_4$ depend on the
specific values of $\alpha_{k}$ and $\alpha_{0}$. From the above constrains, we obtain
\begin{equation}
\frac{p_{C,n}q_{C,n}}{p_{D,n}q_{C,n}}=\frac{p_{C,n}q_{D,n}}{p_{D,n}q_{D,n}}.
\end{equation}
This is a very strong constraint, and generally cannot be satisfied.

\end{appendix}

\end{document}